# Monte Carlo simulation of $1 + 1$-dimensional $\phi^4$ quantum field theory

W. Janke[a]* and T. Sauer[b]

[a]Institut für Physik, Johannes Gutenberg-Universität Mainz, 55099 Mainz, Germany

[b]Institut für Theoretische Physik, Freie Universität Berlin, 14195 Berlin, Germany

We report results of a Monte Carlo simulation of the $\phi^4$ quantum field theory using multigrid simulation techniques and a refined discretization scheme. The resulting accuracy of our data allows for a significant test of an analytical approximation based on a variational ansatz. While the variational approximation is well reproduced for a large range of parameters we find significant deviations for low temperatures and large couplings.

## 1. INTRODUCTION

Among the methods to treat one-dimensional quantum systems analytically the variational approximation has been shown to be a very useful one [1–3]. Since, however, it is hard to give precise inherent error estimates it is desirable to check the variational method against independently obtained numerical data. The difficulty is to achieve sufficient accuracy. Standard path-integral simulations suffer from well-known draw backs, notably systematic errors due to the necessary discretization and severe slowing down in the continuum limit.

In view of these difficulties it is therefore gratifying that recently some algorithmic improvements developed for spin systems and lattice field theories could successfully be transferred to path-integral simulations [4]. Multigrid simulation techniques in particular have been shown to eliminate slowing down in the continuum limit for one-particle systems [5]. Here we will present Monte Carlo (MC) simulation data for the $\phi^4$ quantum chain obtained by combining multigrid techniques with a refined discretization scheme. Finally we take the spatial continuum limit of this system.

## 2. THE MODEL

The $\phi^4$ quantum field theory in its spatially discretized version is a system of $N$ harmonically coupled anharmonic oscillators described by the partition function

$$\mathcal{Z} = e^{-\beta F} = \prod_{i=1}^{N} \int \mathcal{D}[\{\phi_i(u)\}] e^{-\mathcal{H}[\{\phi_i(u)\}]/\hbar}, \quad (1)$$

with a Hamiltonian

$$\mathcal{H} = \int_0^{\hbar\beta} du \left[ Aa \sum_{i=1}^{N} \frac{1}{2}\dot{\phi}_i^2(u) + V(\{\phi_i(u)\}) \right], \quad (2)$$

where the potential is given by

$$V = Aa \sum_{i=1}^{N} \left[ \frac{\omega_0^2}{2}(\phi_i - \phi_{i-1})^2 + \frac{\omega_1^2}{8}(\phi_i^2 - 1)^2 \right]. (3)$$

Here $\beta = 1/k_B T$ denotes the inverse temperature, $\dot{\phi}_i \equiv d\phi_i/du$, and $\mathcal{D}[\{\phi_i(u)\}]$ is the usual path-integral measure. We always employed periodic boundary conditions, $\phi_0 \equiv \phi_N$. Following the notation of Ref.[3] we introduce a dimensionless coupling constant $Q = \hbar\omega_1/E_s$ and parameter $R = \omega_0/\omega_1$, where $E_s = (2/3)Aa\omega_0\omega_1$ is the energy of the classical static kink and $R$ its length in units of the lattice spacing $a$. The reduced temperature will be denoted by $t \equiv k_B T/E_s$.

The variational approach [1–3] starts from a quadratic trial Hamiltonian with free parameters which are determined by optimizing the Jensen-Peierls inequality for the free energy. Already for one-dimensional quantum systems [1,3], however, the resulting set of coupled equations is extremely complicated and explicit solutions have been obtained in the literature only for the limiting cases of high and low temperatures and for small

*W.J. thanks the Deutsche Forschungsgemeinschaft for a Heisenberg fellowship.

coupling $Q$. As a result one obtains an effective classical partition function which can be evaluated, e.g., by standard transfer integral techniques. The small coupling limit, given in eq.(3.6) of Ref.[3], seems to be the most useful one and will be used for comparison with our MC data for the internal energy $u$ per site and the specific heat $c$ per site. More precisely, we will be interested only in the anharmonic contribution $Rdu$ resp. $Rdc$ to these quantities.

## 3. SIMULATION TECHNIQUES

For the numerical work the partition function (1)-(3) was further discretized in euclidean time direction using the Takahashi-Imada scheme [6]. This guarantees a convergence of order $\epsilon^4$, where $\epsilon \equiv \beta/L$ and $L$ is the Trotter number. For the evaluation of the anharmonic contributions the discretization error was further reduced by subtracting the exact values for the harmonic contribution at finite Trotter number $L$ which can readily be derived also for the Takahashi-Imada scheme [8].

Since for local update algorithms we expect a quadratic slowing down in the continuum limit of small $\epsilon$ [5] we applied at each site a multigrid W-cycle with piecewise constant interpolation along the Trotter direction using single-hit Metropolis updating with one presweep and no postsweep. The Metropolis acceptance rates were adjusted to be $\approx 40-60\%$ on the finest grid and the same step width was used for all multigrid levels.

For various $Q$ and $t$ we measured the internal energy using an optimal combination of the "kinetic" and "virial" energy estimators [7,8] with measurements taken every second sweep in runs of 400 000 sweeps after discarding 2 000 sweeps for thermalization. The specific heat was measured by simple numerical differentiation of our combined estimator. All statistical errors were computed by jackkniving the data on the basis of 500 blocks. The energy measurements are practically uncorrelated since the integrated autocorrelation time for both estimators was always bounded by $\tau^{\text{int}} \leq 4$. Within this bound $\tau^{\text{int}}$ tended to be larger for low $t$ and large $Q$.

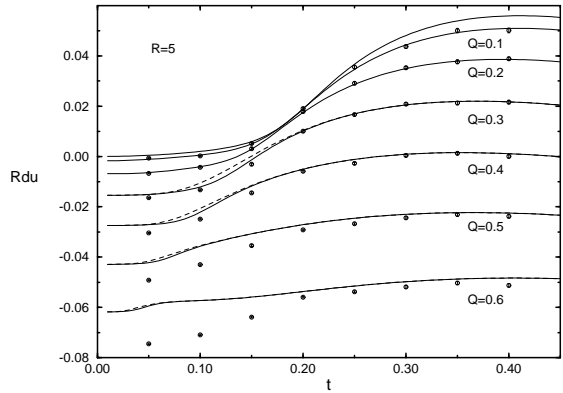

Figure 1. Anharmonic contributions to the internal energy. Solid (dashed) lines show the variational approximation for $N = 300$ ($N = \infty$).

## 4. RESULTS

Figure 1 shows the measured anharmonic contributions to the internal energy. Here the number of oscillators was $N = 300$ except for $t = 0.05$, 0.30, 0.35, and 0.40 where we simulated a chain of $N = 200$ oscillators. The Trotter number was $L = 16$ for $t \geq 0.20$ and $L = 32, 64, 128$ for $t = 0.15, 0.10, 0.05$. By performing a number of simulations with smaller Trotter numbers we convinced ourselves that these values suffice to reduce the remaining discretization to be at most of the order of the statistical error.

Comparing our data with the variational analysis we find that the approximation is fully confirmed for high temperatures $t$ and small couplings $Q$. For lower $t$ we still find a satisfactory agreement if we also take into account finite-size corrections. The situation is different, however, for low temperatures and large couplings where we observe significant deviations from the variational approximation. Note that the error bars for the data in Fig. 1 are smaller than the data symbols. For the worst case, $Q = 0.6$ and $t = 0.05$, the variational approximation deviates from the MC results by 56 statistical error bars, even if finite-size corrections are fully taken into account.

Figure 2 shows the measured anharmonic contributions to the specific heat per site. Since the estimation of the specific heat involves a difference of statistically fluctuating variables the resulting statistical accuracy is greatly reduced compared to the estimation of energies. Therefore our data for the specific heat do not allow for a significant falsifying test of the variational approximation, in particular for large couplings and low temperatures.

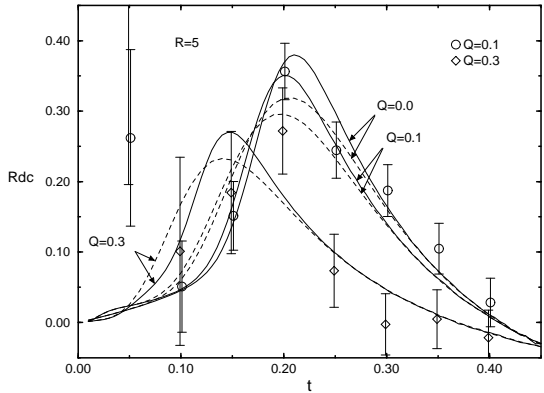

Figure 2. Anharmonic contributions to the specific heat per site. Solid (dashed) lines show the variational approximation for $N = 300$ ($N = \infty$).

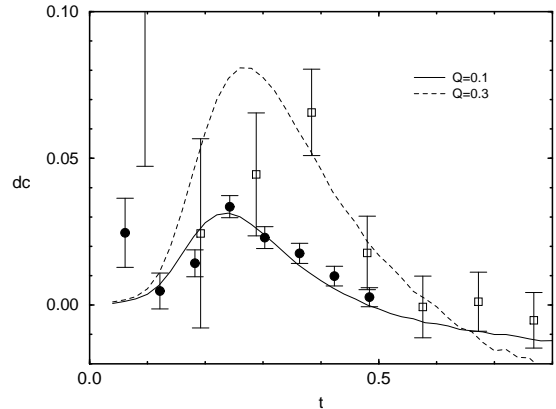

Figure 3. Anharmonic contributions to the specific heat per site of Fig. 2 rescaled to the continuum field theory.

Let us finally consider the renormalized continuum theory. Following the procedure in Ref.[9] we have rescaled in Fig. 3 our data in order to be able to compare them with the variational approximation for the continuum case, i.e., the $\phi^4$ quantum field theory in $1 + 1$ dimensions. For small couplings we find again good agreement between the variational approach and our simulation data.

## 5. DISCUSSION

Employing refined path integral MC techniques to the $\phi^4$ quantum chain we find very good agreement with a variational approximation for small couplings and high temperatures. For large couplings and low temperatures, on the other hand, we observe clear discrepancies which can be caused either by the small coupling expansion used to approximate the effective classical partition function or by an inherent failure of the variational approximation itself (at this level of accuracy). It would therefore be interesting to investigate if higher-order corrections in the coupling parameter $Q$ or higher-order corrections to the variational approach itself can account for the remaining discrepancies.